\def\half{\frac{1}{2}}

\def\bey{\begin{eqnarray}}
\def\eey{\end{eqnarray}}
\def\be{\begin{equation}}
\def\ee{\end{equation}}
\def\ba{\begin{array}}
\def\ea{\end{array}}
\def\gm{\gamma}

\def\bt{\beta}
\def\vep{\varepsilon}

\def\dt{\delta}

\def\pp{\partial}
\def\pp{\partial}

\def\nnb{\nonumber}
\bigskip
\documentclass[onecolumn,showpacs,preprintnumbers,amsmath,amssymb]{revtex4}
\usepackage[colorlinks,urlcolor=blue,citecolor=blue,linkcolor=blue]{hyperref}
\usepackage{graphicx}
\usepackage{fancyhdr}
\usepackage{dcolumn}
\usepackage{bm}
\usepackage{multirow}
\usepackage{amssymb}
\usepackage{amsmath}
\usepackage{graphicx}
\usepackage[normalem]{ulem}
\usepackage[dvips]{color}

\begin{document}
\preprint{ }
\title{ Effects of fermionic dark matter on properties of neutron stars}
\author{Qian-Fei Xiang$^{1,2}$,
 Wei-Zhou Jiang$^{1,3,4}$\footnote{wzjiang@seu.edu.cn}, Dong-Rui Zhang$^1$, and Rong-Yao Yang$^1$}
 \affiliation{  $^1$ Department of Physics, Southeast University,
Nanjing 211189, China\\
$^2$Institute of High Energy Physics, Chinese Academy of Sciences, Beijing 100049, China\\
$^3$ National Laboratory of Heavy Ion Accelerator, Lanzhou 730000,
China\\
 $^4$ Department of Physics and Santa Cruz Institute for Particle Physics, University of California,
 Santa Cruz, CA 95064, USA}

\begin{abstract}
\baselineskip18pt By assuming that only gravitation exists between
dark matter (DM) and normal matter (NM), we study the effects of
fermionic DM on the properties of neutron stars using the two-fluid
Tolman-Oppenheimer-Volkoff formalism. It is found that the
mass-radius relationship of the DM admixed neutron stars (DANSs)
depends sensitively on the mass of DM candidates, the amount of DM,
and interactions among DM candidates. The existence of DM in DANSs
results in a spread of mass-radius relationships that cannot be
interpreted with a unique equilibrium sequence.  In some cases, the
DM distribution can surpass the NM distribution to form DM halo. In
particular, it is favorable to form an explicit DM halo, provided the
repulsion of DM exists.  It is interesting to find that the
difference in particle number density distributions in DANSs and
consequently in star radii caused by various density dependencies of
nuclear symmetry energy tends to disappear as long as the repulsion
of accumulated DM is sufficient. These phenomena indicate that the
admixture of DM in neutron stars can significantly affect the
astrophysical extraction of nuclear equation of state by virtue of
neutron star measurements. In addition, the effect of the DM
admixture on the star maximum mass is also investigated.

\end{abstract}
 \keywords{} \maketitle
\baselineskip 20.6pt

\section{Introduction}
Nowadays, dark matter (DM) becomes a very hot topic in both
astrophysics~\cite{Roos} and particle physics~\cite{Feng}, and it
seems clear that it is the  dominant matter in the universe. Recent
advances in cosmological precision tests further consolidate the
minimal cosmological standard model, indicating that the universe
contains 4.9\% ordinary matter, 26.8\% DM, and 68.3\% dark
energy~\cite{Komatsu}. However, the properties of DM, including its
mass and  interactions, are still unknown. It is thus of great
interest to explore  the properties of DM through direct or indirect
methods.

There are usually three well-known methods to detect DM particles:
using particle accelerators~\cite{Arcadi}, direct detecting of
scattering cross section by terrestrial detectors (CDMS II, CRESST,
and CoGeNT ), and indirect detecting of products from DM particle
annihilation in the galactic halo~\cite{Feng2}. The latest
experimental results are not conclusive. The data from
DAMA/LIBRA~\cite{DAMA}, CoGeNT~\cite{Co1,Co2}, and CRESST-II
~\cite{cresst2} experiments may imply the light DM particle with the
mass around $8-10 GeV$, while other experiments such as
CDMS-II~\cite{CDMS2}, SIMPLE~\cite{SIMPLE} and
XENON$10/100$~\cite{XENON10,XENON100} reported null results at the
same region. It's suggested that isospin-violating DM  would relax
the tensions between the results of DAMA, CoGeNT and XENON
experiments~\cite{Jin}. But the possible tensions between some
experiments such as those between DAMA and SIMPLE are unlikely to be
affected by isospin-violating interactions.

One indirect method that is gaining attention in recent years is to
study the DM effects on compact stars. On one hand, the large
baryonic density in compact stars increases the probability of DM
capture within the star and eventually results in gravitational
trapping~\cite{gold89,kouv08}. The DM accumulation inside stars will
affect the stellar structure and even contribute to the collapse of a
neutron star~\cite{kouv11,kouva11,Kouvaris,Lavallaz}.  On the other
hand, at the later evolution period, neutron stars can be rather cold
due to lack of possible burning or heating mechanisms, and therefore
the heating effect by possible DM annihilation  is favorable and
possibly observable\cite{Lavallaz}. At the same time,
self-annihilation of DM in the inner regions of neutron stars may
significantly affect their kinematical properties, including linear
and angular momentum~\cite{Angeles}. Therefore, it is of great
significance to study the  potential effects of DM on the properties
of neutron stars.

The neutron star properties  are intimately related to the nuclear
equation of state (EOS) of asymmetric matter, while the latter
consists roughly of the EOS of symmetric matter and the density
dependence of the nuclear symmetry  energy. In the past, great
progress has been achieved to constrain the EOS of symmetric nuclear
matter using terrestrial nuclear experiments for more than three
decades, see, e.g.~\cite{You99,da02}, for a review. It is known that
the EOS around normal density can be well constrained by nuclear
giant monopole resonances~\cite{You99}, and at supra-normal densities
it can be constrained by the collective flow data from high energy
heavy-ion reactions~\cite{da02}. Stringent constraints on the nuclear
EOS are indispensable to portray neutron stars.  For instance, the
maximum mass of neutron stars is mainly determined by the
high-density EOS of symmetric nuclear matter. Recently, several
neutron stars with large masses around $2M_\odot$ have been observed
~\cite{Ni05,Ni08,Oz06,De10,An13}. In particular, the $2M_\odot$
pulsar J1614-2230 was measured rather accurately through the Shapiro
delay~\cite{De10}, and it was reported most recently that the
accurate measurement of $2M_\odot$ pulsar J0348+0432 would be more
restrictive~\cite{An13}. These observations can provide astrophysical
constraints on the high-density nuclear EOS. On the other hand, the
density dependence of the symmetry energy that affects mostly the
radius of NS's is still not well determined especially at high
densities, though in the past great endeavors have been made to
constrain it in the terrestrial laboratories, e.g. see
Ref.~\cite{Li08}. Since neutron stars are natural laboratories for
the exploration of baryonic matter under extreme conditions,  it is
hopeful to constrain the nuclear EOS of asymmetric matter at
densities of interest jointly using the radius and mass observations
of neutron stars~\cite{St2010,Oz09}. One should, however, note that
such constraints can not provide detailed information of the neutron
star compositions or interplay between them. As far as the DM is
concerned, it is rather interesting to explore many relevant issues
of DM, such as its accretion onto neutron stars, interplay between
normal matter (NM) and DM, interactions among DM, its effect on
properties of neutron stars, and so on.

The dark-matter admixed neutron stars (DANSs) have been studied
recently in several articles~\cite{Li, Sandin1, Sandin, Leung,
Kouvaris1, Lavallaz}. Regarding DM as Fermi-gas-like matter, Li et.
al.~\cite{Li} provided an equal admixture of DM and NM in DANSs by
taking the total pressure (energy density) as the simple sum of those
of the DM and NM. Sandin and Ciarcelluti~\cite{Sandin1,Sandin}
considered the mirror DM and assumed that neutron stars with a DM
core are inherently two-fluid systems where the NM and DM couple
essentially only through gravity. By varying the relative size of the
DM core, they may reproduce all astrophysical mass measurements based
on one nuclear matter EOS. Leung et al.~\cite{Leung} used the general
relativistic two-fluid formalism to study the various structures of
DANS. Following the approach used by Sandin and
Ciarcelluti~\cite{Sandin1,Sandin},  we consider in this work a broad
variety of DM with various masses and interactions in neutron stars
and examine the effects of DM on static properties of neutron stars.
The paper is organized as follows. In Sec. II, we present briefly the
formalism of DM based on the Lagrangian of the relativistic
mean-field models as well as the formalism of two-fluid model for
neutron stars.  In Sec. III, numerical results and discussions are
presented.
 At last, a summary is given in Sec. IV.

\section{Formalism}
\label{rmf} In the RMF approach, the nucleon-nucleon interaction is
usually described via the exchange of three mesons: the isoscalar
meson $\sigma$, which provides the medium-range attraction between
the nucleons, the isoscalar-vector meson $\omega$, which offers the
short-range repulsion, and the isovector-vector meson $b_0$, which
accounts for the isospin dependence of the nuclear force. The
relativistic Lagrangian can be written as: \bey
 {\cal L}&=&
{\overline\psi}[i\gm_{\mu}\partial^{\mu}-M+g_{\sigma}\sigma-g_{\omega}
\gm_{\mu}\omega^{\mu}-g_\rho\gm_\mu \tau_3 b_0^\mu
   ]\psi\nnb\\
      &  &
    - \frac{1}{4}F_{\mu\nu}F^{\mu\nu}+
      \frac{1}{2}m_{\omega}^{2}\omega_{\mu}\omega^{\mu}
    - \frac{1}{4}B_{\mu\nu} B^{\mu\nu}
     +\frac{1}{2}m_{\rho}^{2} b_{0\mu} b_0^{\mu}\nnb\\
     &   &
+\frac{1}{2}(\partial_{\mu}\sigma\partial^{\mu}\sigma-m_{\sigma}^{2}\sigma^{2})
+U_{\rm eff}(\sigma,\omega^\mu, b_0^\mu). \label{eq:lag1}
  \eey
 where
 $\psi,\sigma,\omega$,$b_0$ are the fields of
the nucleon, isoscalar, isoscalar-vector, and neutral
isovector-vector mesons, with their masses $M, m_\sigma,m_\omega$,
and $m_\rho$, respectively. $g_i(i=\sigma,\omega,\rho)$  are the
corresponding meson-nucleon couplings. $F_{\mu\nu}$ and $ B_{\mu\nu}$
are the strength tensors
 of $\omega$ and $\rho$ mesons respectively,
\begin{equation}\label{strength} F_{\mu\nu}=\pp_\mu
\omega_\nu -\pp_\nu \omega_\mu,\hbox{  } B_{\mu\nu}=\pp_\mu b_{0\nu}
-\pp_\nu b_{0\mu}.
\end{equation}
The self-interacting terms of $\sigma$, $\omega$ mesons and  the
isoscalar-isovector coupling  are given generally as
 \begin{equation}
 U_{\rm eff}(\sigma,\omega^\mu, b_0^\mu)=-\frac{1}{3}g_2\sigma^3-\frac{1}{4}g_3\sigma^4
 +\frac{1}{4}c_3(\omega_\mu\omega^\mu)^2
 +4 {\Lambda}_V g_{\rho}^2 g_{\omega}^2
 {\omega}_{\mu}{\omega}^{\mu} b_{0 \nu} b^{0 \nu}.
 \label{eq:u}
 \end{equation}
With these nonlinear meson self-interaction terms, the models are
usually known as the nonlinear models. In the relativistic mean-field
(RMF) approximation, the energy density $\vep$ and pressure $p$ are
written as:
 \bey
 {\vep}&=&\sum_{i=p,n}\frac{2}{(2\pi)^3}\int^{k_{F_i}} d^3\!k E^*+
 \frac{1}{2}m_\omega^2\omega_0^2+
 \frac{1}{2}m_\sigma^2\sigma_0^2+\frac{1}{2}m_\rho^2 b_0^2 \nnb\\
&&+\frac{1}{3}g_2\sigma_0^3+\frac{1}{4}g_3\sigma_0^4+\frac{3}{4}c_3\omega_0^4
+12 {\Lambda}_V g_{\rho}^2 g_{\omega}^2{\omega}_0^2 b_0^2,\label{eq:e}\\
p&=&\frac{1}{3}\sum_{i=p,n}\frac{2}{(2\pi)^3}\int^{k_{F_i}} d^3\!k
\frac{{\bf k}^2}{E^*}+\frac{1}{2}m_\omega^2\omega_0^2
-\frac{1}{2}m_\sigma^2\sigma_0^2+\frac{1}{2}m_\rho^2
b_0^2\nnb\\
&&-\frac{1}{3}g_2\sigma_0^3-\frac{1}{4}g_3\sigma_0^4+\frac{1}{4}c_3\omega_0^4
+4 {\Lambda}_V g_{\rho}^2 g_{\omega}^2{\omega}_0^2 b_0^2,
\label{eq:p}\eey where $\sigma_0$, $\omega_0$, and $b_0$ are the
scalar, vector, and isovector-vector fields in the mean-field
approximation, respectively,  $E^*=\sqrt{{\bf k}^2+(M^*)^2}$, and
$k_F$ is the Fermi momentum.

In this work, we also invoke density-dependent RMF models SLC and
SLCd proposed by Jiang et. al. based on the Brown-Rho
scalings~\cite{ji07,ji07b}. In these density-dependent RMF models,
there are no nonlinear meson self-interaction terms, and the energy
density and pressure are given as
 \bey
 {\vep}&=&\half C_\omega^2\rho^2+\half C_\rho^2 \rho^2\dt^2+ \half
 \tilde{C}^2_\sigma(M-M^*)^2 +\sum_{i=p,n}
 \frac{2}{(2\pi)^3}\int_{0}^{{k_F}_i}\! d^3\!k~ E^*,  \label{eqe1} \\
 p&=&\half C_\omega^2\rho^2+\half C_\rho^2 \rho^2\dt^2- \half
 \tilde{C}^2_\sigma(M-M^*)^2 -\Sigma_0\rho+
 \frac{1}{3}\sum_{i=p,n}\frac{2}{(2\pi)^3}\int_{0}^{{k_F}_i}\! d^3\!k
 ~\frac{{\bf k}^2}{E^*}
\label{eqp1}
 \eey
where $C_\omega=g_\omega^*/m_\omega^*$, $C_\rho=g_\rho^*/m_\rho^*$,
$\tilde{C}_\sigma=m_\sigma^*/g_\sigma^*$, $M^*=M-g^*_\sigma\sigma_0$
is the effective mass of nucleon, $\dt=(\rho_n-\rho_p)/\rho$ is the
isospin asymmetry with $\rho$ being the total number density of
nucleons, and $\Sigma_0$ is the rearrangement term due to the density
dependence of the parameters. Here, the meson coupling constants and
hadron masses with asterisks denote the density dependence, given by
the BR scaling.

Given above is the formalism for nuclear matter without considering
the $\beta$ equilibrium. For asymmetric nuclear matter at $\beta$
equilibrium, the chemical equilibrium and charge neutrality
conditions need to be additionally considered, which are written as:
 \bey
  \mu_n&=&\mu_p+\mu_e,\\
  \rho_e&=&\rho_p,\\
  \rho&=&\rho_n+\rho_p.
 \eey
where $\rho_e$ is the number density of electrons, and $\mu_n, \mu_p$
and $\mu_e$ are the chemical potential of neutron, proton and
electron, respectively. For neutron star matter, we need also to
include the contribution of the free electron gas in
Eqs.(\ref{eq:e}), (\ref{eq:p}), (\ref{eqe1}) and (\ref{eqp1}).

We regard DM candidates as fermions, and assume that a neutral scalar
meson couples to the DM candidate through
  \begin{math}g_{\rm s}\bar{\psi_{\rm D}}\psi_{\rm D}\phi\end{math}
   and that a neutral
 vector meson couples to the conserved DM current through
\begin{math}
g_{\rm v}  \bar{\psi_{\rm D}}  \gamma _{\rm \mu}    \psi_{\rm D}
V^{\rm \mu}
\end{math}.
This modelling of the interactions is rather universal according  to
the covariance, and it's an extension of those used in
Ref.~\cite{Narain}. Similar to the potential for
baryons~\cite{Fetter}, the meson exchange gives rise to an effective
potential for DM candidates: \be
       V_{eff}(r)=\frac{g_{v}^{2}}{4\pi} \frac{e^{-m_{v}r}}{r}
       -\frac{g_{s}^{2}}{4\pi} \frac{e^{-m_{s}r}}{r}.
       \label{potential}
\ee With appropriate coupling constants and masses, the above
potential is attractive at large separations and repulsive at short
distances.

The Lagrangian density for the present DM model can be written as:
 \bey
 {\cal L_D}&=&
{\overline\psi_D}[\gamma_\mu (i\partial^\mu-g_\mu V^\mu)-(M_D -g_s \phi)]\psi_D
+\frac{1}{2}(\partial_{\mu}\phi \partial^{\mu}\phi-m_{s}^{2}\phi^{2})  \nnb\\
&& - \frac{1}{4}D_{\mu\nu} D^{\mu\nu}+\frac{1}{2}m_{v}^{2} V_{\mu}
V^{\mu}
  \eey
where $D_{\mu\nu}$ is the strength tensor of $vector$ meson $$
D_{\mu\nu}= \partial_{\mu} V_{\nu}-\partial_{\nu} V_{\mu}.$$

The relativistic quantum field theory generated by this
Lorentz-invariant ${\cal L_D}$ is renormalizable, for it is similar
to the massive QED with a conserved current and an additional scalar
interaction. The energy density and pressure of DM are given as:
 \bey
 \vep_D &=&  \frac{2}{(2\pi)^3}\int^{k_{F_D}} d^3\!k \sqrt{{\bf k}^2+(M_D^*)^2}
 +\frac{g_\nu^2}{2 m_\nu^2}\rho_D^2+
 \frac{m_s^2}{2 g_s^2} (M_D-M_D ^*)^2,\label{eq:ed}
 \\
p_D&=&\frac{1}{3}\frac{2}{(2\pi)^3}\int^{k_{F_D}} d^3\!k
 \frac{{\bf k}^2}{\sqrt{{\bf k}^2+(M_D^*)^2}}
+\frac{g_\nu^2}{2 m_\nu^2}\rho_D^2-
 \frac{m_s^2}{2 g_s^2} (M_D-M_D ^*)^2,\label{eq:pd}
 \eey
where $\rho_D $ is the number density of DM, $M_D$ is the mass of the
DM candidate, $M_D^*$ is the effective mass of DM determined by
$M_D^* = M_D - g_s\phi_0$ with $\phi_0$ being the RMF scalar field.

To study properties of the DANS, we necessarily adopt the two-fluid
formulism for NM and DM between which only the gravitation exists.
For a static and spherically symmetric space-time $d\tau^2 =
e^{2\nu(r)}dt^2-e^{2\lambda(r)}dr^2-r^2(d\theta^2+\sin^2\theta
d\phi^2)$. The two-fluid Tolman-Oppenheimer-Volkoff (TOV) equations
for the DANS are given as~\cite{Sandin} (units are chosen such that
G$=$ c $=$ 1) \bey
\frac{d \nu}{dr} = \frac{M(r)+4\pi r^3 p(r)}{r[r-2M(r)]},\nnb\\
\frac{d p_N}{dr} = -[p_N(r)+\vep_N(r)]\frac{d\nu}{dr},\nnb \\
\frac{d p_D}{dr} = -[p_D(r)+\vep_D(r)]\frac{d\nu}{dr},
\label{eq:TOVS} \eey where $r$ is the radial coordinate from the
center of the star, $p_N(r)$ [$p_D(r)$] and $\vep_N(r)$
[$\vep_D(r)$]are the pressure and energy density of NM [DM] at
position $r$. $p(r)=p_N(r)+p_D(r)$ is the sum of the pressures at the
position $r$, while $M(r)=\int^{r} d\tilde r 4\pi {\tilde
r}^2(\vep_N(\tilde r)+\vep_D(\tilde r))$ is the sum of masses
contained in the sphere of the radius $r$. Here, the separate
relations for the pressure of normal and dark matter accord with the
assumption that there is no interaction between normal and dark
matter except the gravitational interaction that modifies the metric
together. Specifically, Eq.(\ref{eq:TOVS}) can be deduced from the
stationary condition of the star mass, that is, the variation of
total star mass should vanish with respect to the variations of the
energy and particle densities of normal and dark matter. The detailed
derivation of Eq.(\ref{eq:TOVS}) is given in Appendix~\ref{append}.

The total radius $R$ and mass $M(R)$ of a DANS are determined by the
condition $p(R)=0$.  On the other hand,  we can obtain the radius and
mass individually for DM and NM according to the conditions $p_N
(R_N)=0$ and $p_D (R_D)=0$. Because NM in a DANS, consisting of
neutrons, protons, and electrons (npe) at $\bt$ equilibrium in this
work, undergoes a phase transition from the homogeneous matter to the
inhomogeneous matter at the low density region, the RMF EOS obtained
from the homogeneous matter can not apply to the low density region.
For a thorough description of the NM part in DANSs, we thus adopt the
empirical low-density EOS in the literature~\cite{Ba71,Ii97}, while
for the DM part, no similar treatment is considered at the surface.

\section{Numerical results and discussions}
\label{results}

In the universe, DM has been proposed to explain the mass
discrepancies according to the observed galactic rotation velocities.
The latest data indicate that DM occupies about 26.8\% of the total,
while the ordinary visible matter only has the proportion of 4.9\%
and the leftover is the Dark energy. For DM as the majority of
matter, people have gained few information of its mass and
interaction strength. Except for the gravitational effect, the
astrophysical objects composed of DM are regarded to be too faint to
be detectable. DM can be possibly accreted in neutron stars by losing
its energy through repeated scatterings in the high-density medium,
while the practical amount of dark matter accreted onto normal
neutron stars should depend on the dark matter-baryon cross section
and the age of the neutron star~\cite{gold89,kouv08}. Probably, the
accretion of DM can become easier if the DM that already resides in
stars has necessary self-interaction. On the other hand, since the
majority of matter is DM in our universe, one can imagine the
existence of a large number of the massive objects composed of DM in
our universe. Similar accretion of NM can take place in those DM
stars.  In the present work, we thus assume the DM fraction in
neutron stars as an arbitrary quantity, regardless of accretion
details.  Similar to that in Ref.~\cite{Narain}, here we consider DM
candidates with various masses and interaction strengthes in compact
stars. Though the present consideration turns out to be somehow
unconstrained, we are actually interested in possible constraints for
DM by examining the its effects on the neutron star radius and mass,
including the maximum mass.

For NM, we adopt the nuclear EOS with the density-dependent RMF
models SLC and SLCd~\cite{ji07,ji07b}. Because the unique difference
between the models SLC and SLCd is that the latter has a softer
symmetry energy than the former, we may investigate the symmetry
energy effect on neutron stars involving the DM. For comparisons, we
also perform calculations with the nonlinear RMF model
IU-FSU~\cite{IUFSUG}. Parameters and saturation properties of these
RMF models are listed in Table~\ref{t:t1}.

\begin{table}[h]
\caption{Parameters and saturation properties for various RMF models.
For the density-dependent models SLC and SLCd, tabulated are the
parameters at zero density, while the parameters scaling the density
dependence can be found in Ref.~\cite{ji07b} and are not listed here.
Meson masses, incompressibility and symmetry energy are in units of
MeV, and the density is in unit of $fm^{-3}$. \label{t:t1}}
 \begin{center}
    \begin{tabular}{ c c c c c c c c c c c c c  c}
\hline\hline &$g_\sigma$ & $g_\omega$& $g_\rho$& $m_\sigma$ &
$m_\omega$ & $m_\rho$ & $g_2$ & $g_3$ & $c_3$ & $\Lambda_v$& $\rho_0$
&
$\kappa$ & $E_{sym}$ \\
\hline
SLC     &10.141 &10.326 &3.802 &590.000 &783.000 &770 &-      &- &-      &- &0.16 &230 &31.6\\
SLCd    &10.141 &10.326 &5.776  &590.000 &783.000 &770 &-      &-  &-      &- &0.16 &230 &31.6\\
IU-FSU     &9.971 &13.032 &6.795 &508.194 &782.501 &763 &8.493 &0.488 &144.219 &0.046  & 0.155 &231.2 &31.3\\
\hline\hline
\end{tabular}
\end{center}
\end{table}

\begin{figure}[thb]
\begin{center}
\vspace*{-5mm}
\includegraphics[height=8.0cm,width=14.0cm]{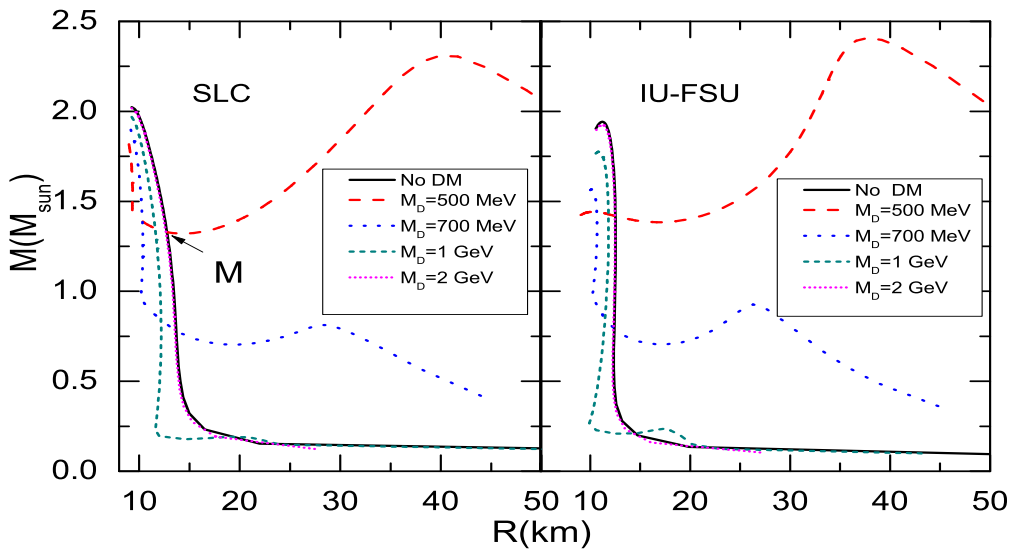}
 \end{center}
\vspace*{-5mm}\caption{(Color online) Mass-radius relations of the
DANSs with the DM candidate mass ($M_D$) ranging from 500MeV to 2GeV.
The results in the left and right panels are obtained with the SLC
and IU-FSU, respectively. \label{f:MASS}}
\end{figure}

Here, the central energy density of DM and NM is assumed to be the
same. Yet, we have no preference to use this unitary ratio at the
neutron star center. Later on, we will come back to this point to see
the effect by changing this ratio. Even if the unitary ratio at the
neutron star center is used, the actual admixture of DM and NM  in
DANSs is determined by the two-fluid TOV equations and thus is not
equal at all. Shown in Fig.~\ref{f:MASS}  are the Mass-radius
relations for two models  with the inclusion of different kinds of
free DM particles that differ in masses.  It is seen clearly that if
the mass of the DM candidate is much larger, e.g.$\geq 2 GeV$, DM has
little impact on the Mass-radius relations of DANSs. If the mass of
DM is small (e.g. $500MeV$), it can influence the mass-radius
relations of DANS significantly. In this case,  the minimum mass we
obtain is found to be too heavy ($M \geq 1.32 M\odot$ for SLC and $M
\geq 1.38 M\odot$ for IU-FSU) to explain the low-mass neutron stars
observed~\cite{littimer,ker,tho}. However, if we relax the condition
that the central energy densities are the same for  NM and DM by
reducing the central energy density of DM, we can obtain a much
smaller minimum mass to be consistent with the observation of the
low-mass neutron stars. By now, neither can we confirm the DM
candidate terrestrially, nor can we know the mass of DM candidates.
There could be  many choices for DM candidates from lightest axions
($M\sim \mu eV-10meV$) to heaviest WIMPs
($M\sim10GeV-TeV$)~\cite{Feng}. Our research implies that  one can
perhaps distinguish them through their impact on neutron stars.
Recently, an analysis about the XENON10 data claimed that direct
detection experiments can be sensitive to DM candidates with masses
well below the GeV scale~\cite{XENON10}. Our calculation indicates
that  the observation of neutron stars is also sensitive to the DM
admixture with the DM candidate mass below the GeV scale. Note that
the identification of DM with the neutron star observations also
involves the details of the celestial DM distribution and accretion.

\begin{figure}[thb]
\begin{center}
\vspace*{-5mm}
\includegraphics[height=8.0cm,width=11.0cm]{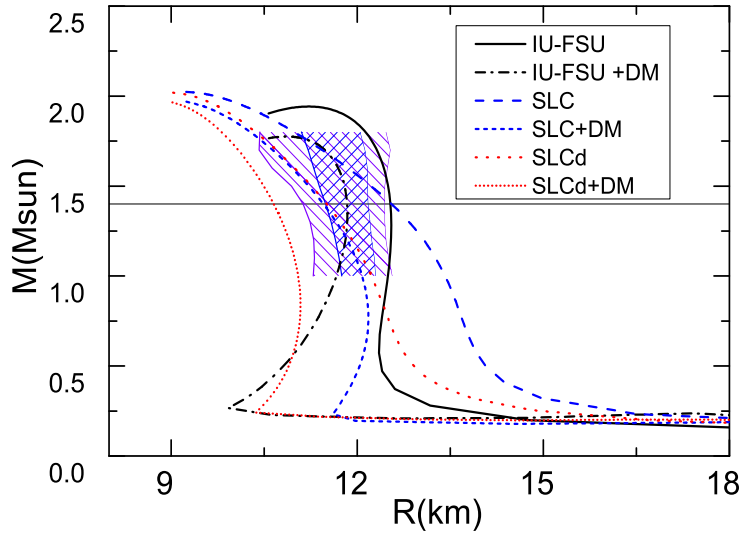}
 \end{center}
\vspace*{-5mm}\caption{(Color online) Mass-radius relations of
neutron stars and DANSs for different models. In the legend, the RMF
models without the suffix DM represent results of  normal neutron
stars, while the ones with the suffix DM stands for results of
DANSs. The hatched areas give the probability distributions with
$1\sigma$ (blue) and $2\sigma$ (violet) confidence limits
 for $ r_{ph}>>R $ summarized in Ref.~\cite{St2010}.\label{f:MR}}
\end{figure}

In Fig.~\ref{f:MR}, it shows the mass-radius relations of DANSs for
different models. Here, the mass of free DM candidate is taken to be
1 GeV. We see that the existence of  DM would lower the maximum
stable mass of neutron stars allowed by the corresponding EOS. The
reduction of the maximum mass due to the inclusion  of DM is
different for different models. We see that the reduction is indeed
not apparent with the models SLC and SLCd that feature a stiff EOS at
high densities~\cite{ji07}. As shown in Fig.~\ref{f:MR}, more
apparent phenomenon occurs for the shift of the neutron star radius
caused by the inclusion of DM. For instance, by comparing with the
neutron star and DANS that both have the mass $1.4M\odot$, denoted by
the straight line in Fig.~\ref{f:MR}, we see that very significant
reduction of the radius arises with the inclusion of DM in neutron
stars.  It is of special interest to point out that the 1GeV DM
candidate is possibly relevant to the mirror particle of visible
nucleonic matter~\cite{Sandin,bli83,khl91}. In Fig.~\ref{f:MR}, we
also include the constraints of mass-radius for $r_{ph} >>R$ obtained
by Steiner et al.\cite{St2010}. We see that the inclusion of DM can
lead to a more vertical shape, which is favorable for models SLC and
SLCd to be consistent with observational constraints. Based on the
general understanding of neutron star properties, the slope of the
nuclear symmetry energy at normal density was extracted to be in the
range 36-55 MeV at the 95\% confidence level~\cite{st2012}. Were DM
to be present in neutron stars, the astrophysical extraction of the
symmetry energy should suffer significant modification due to the
distinct effects induced by DM. On the other hand, though the
inclusion of dark matter favors the vertical shape and smaller
radius, we should point out that the comparison with the constraints
obtained by Steiner et al. does not seem to be direct since such
constraints were extracted using the one-fluid TOV equation.

\begin{figure}[thb]
\begin{center}
\vspace*{-5mm}
\includegraphics[height=8.0cm,width=10.0cm]{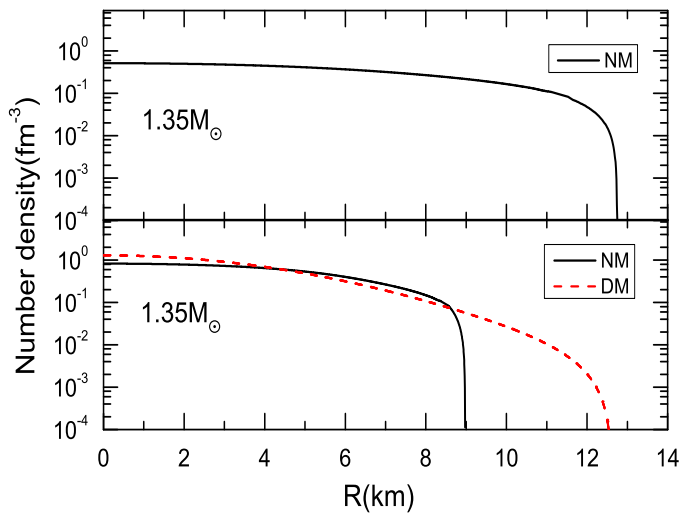}
 \end{center}
\vspace*{-5mm}\caption{(Color online) Density profiles in the neutron
star and DANS that have the same mass $1.35M_\odot$ and radius
(12.76km). For the DANS, the DM candidate mass is 0.5 GeV. The mass
and radius setup corresponds to the point marked by M in
Fig.\ref{f:MASS}. The upper panel is for the normal neutron star, and
the lower panel is for the DANS.\label{NB}}
\end{figure}
Given the appreciable DM effect on the neutron star radius, it is
interesting to compare and examine various matter distributions in
neutron stars and DANS. For this aim, we choose the neutron star and
DANS that have the same mass and radius.   The neutron star marked by
M, as shown in Fig.~\ref{f:MASS}, is the exact one that meets the
criterion. The intersection point M corresponds for the neutron star
and DANS to  the mass ($1.35M_\odot$) and R (12.76km), while here the
mass of DM candidate is 0.5 GeV. The number density profiles with the
SLC are displayed in Fig.~\ref{NB}: the upper panel for the normal
neutron star and the lower panel for the DANS. Significant difference
in the density profiles in the neutron star and DANS can be observed.
In the DANS, we see that a smaller NM core ($R\sim9km$) is surrounded
by a DM halo. Since the two stars have the same mass,  it seems
impossible to distinguish them only by their gravitational effects on
other neighboring stellar objects. However, their visible radii are
different. Thus, it is possible to distinguish them by measuring the
gravitational redshift of spectral lines. In fact, the radius of the
neutron star is 12.76 km, while the visible radius (namely the radius
of the NM core) of the DANS is 9 km. According to the gravitational
redshift formula: $z=\frac{1}{\sqrt{1-2GM/Rc^2}}-1$, the redshift of
the neutron star is obtained to be 0.206, and it is 0.340 for the
DANS.

\begin{figure}[thb]
\begin{center}
\vspace*{-5mm}
\includegraphics[height=8.0cm,width=14.0cm]{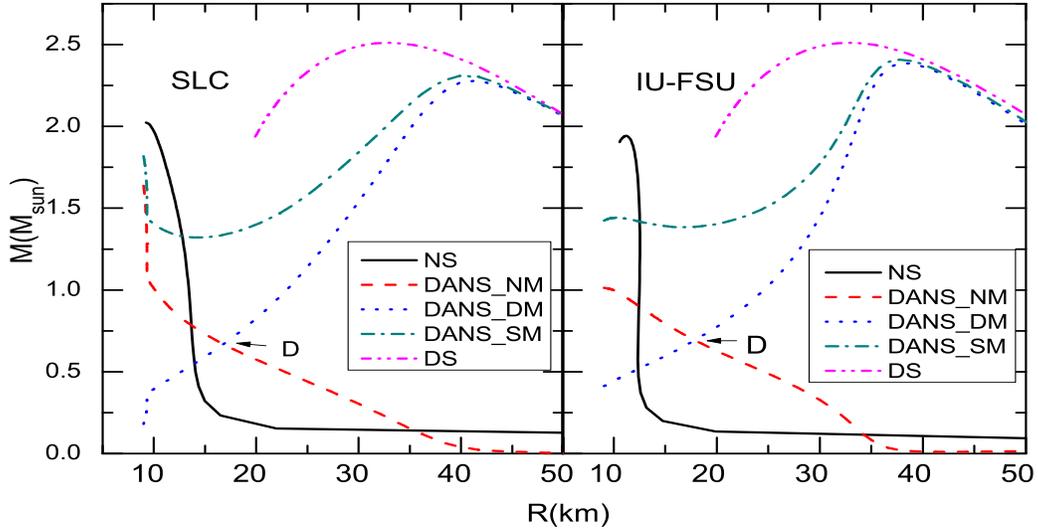}\end{center}
\vspace*{-5mm}\caption{(Color online) Constituent masses in the DANS
with the DM particle mass being 0.5 GeV. DANS$_-$NM (DANS$_-$DM) is
for the mass-radius relation of NM (DM) in DANS, and DANS$_-$SM
corresponds to the total one. Besides, curves for normal neutron star
and pure DM star (DS) are depicted. \label{DMD}}
\end{figure}

In order to further study the behavior shown in Fig.~\ref{f:MASS}, we
depict the mass-radius relations of DM and NM in Fig.~\ref{DMD}.
Here, the mass of DM candidate is again chosen to be 0.5 GeV, and the
ratio of the central DM energy density to that of NM  is fixed to be
equal. It's obvious that with the  increase of the radius, the mass
proportion of NM in DANSs decreases, while the mass proportion of DM
in DANSs increases. The reason for this phenomenon lies in the fact
that in the current two-fluid model DM particles interact with NM
particles only through the gravity. Both  DM and NM produce their own
maximum masses and corresponding radii. For neutron stars without DM,
the maximum mass  is $2.02 M_\odot$ ($1.94 M_\odot$) and the
corresponding radius is  9.24km (11.21km) with the SLC (IU-FSU). With
the inclusion of DM, the original maximum mass with the SLC (IU-FSU)
is reduced to  $1.82 M_\odot$ ($1.44 M_\odot$), and corresponding
radius is decreased to 8.98km (10.3km). At the same time, for pure DM
star, the maximum mass is determined by $M_{max} = 0.627 M_\odot
(\frac{1GeV}{M_D})^2$ and corresponding radius  is $R_{D}
=8.115km(\frac{1GeV}{M_D})^2 $~\cite{Narain}. By substituting $M_D=
0.5GeV$, we get the maximum mass  2.5 $M_\odot$ and corresponding
radius  32.46km.  When the DM admixture starts to take place, the
maximum mass and corresponding radius will be changed, as shown in
Fig.~\ref{DMD}. At the cross point (marked by D), the mass
proportions of DM and NM in the DANS are equal. As the star radius
continues to increase, the mass proportion of DM will be larger than
that of NM, eventually leading to  DM-dominant stars.
\begin{figure}[thb]
\begin{center}
\vspace*{-5mm}
\includegraphics[height=8.0cm,width=14.0cm]{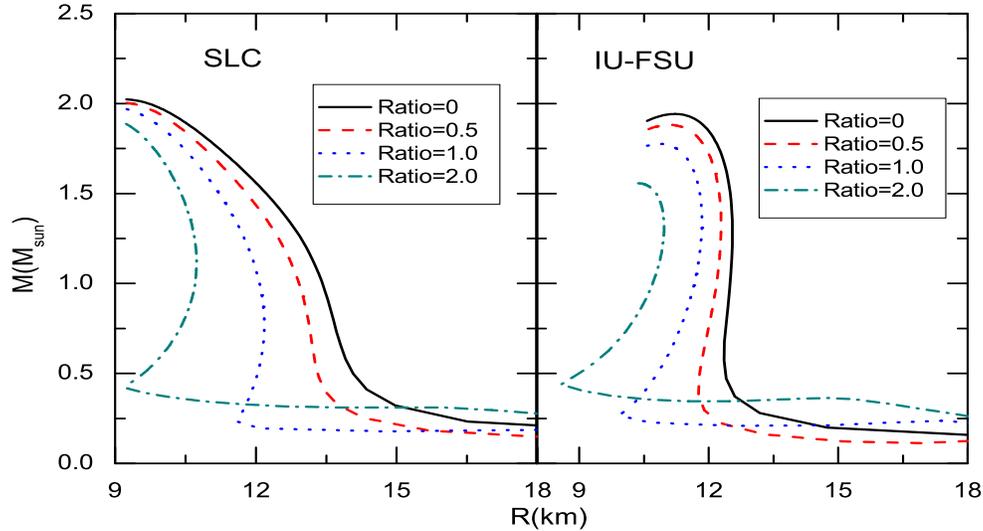}
 \end{center}
\vspace*{-5mm}\caption{(Color online) Mass-radius relations for
various DM proportions.  The  Ratio parameter is the central energy
density ratio of the DM to the NM. The zero ratio means DM-free
normal neutron star.\label{RATIO}}
\end{figure}

Shown in Fig.~\ref{RATIO} are the mass-radius relations of DANSs for
various proportions of DM. Here, the mass of the DM candidate is
taken to be $M_D=1GeV$. We see that with increasing the DM ratio, the
maximum mass and corresponding radius of DANS decrease.  At the same
time, the minimum mass of DANS increases. Similar to results shown in
previous figures, the reduction of maximum mass in SLC is much less
appreciable than that in IU-FSU. It is seen that the structure of
DANSs is dependent on the amount of DM, and this results in a spread
of mass-radius relations that cannot be interpreted with a unique
equilibrium sequence.

Generally speaking, the amount of DM in DANSs is expected to vary and
depends on the whole history of the star~\cite{gold89,kouv08},
especially on the environment from which it originates and in which
it lives. In fact, the capture of DM is difficult both because of
their low reaction cross section (typically about $10^{-44}
cm^2$~\cite{ahmed}) and the low average density of matter in the
universe. But NM galaxy and galaxy clusters could act as a whole to
capture DM, and enable the density of DM in some section to be high
enough to produce observable signals. Actually, most evidences of DM
come from these observations~\cite{refregier,Tyson,Lewis}. Recently,
a research claimed that neutron stars in binary systems might
increase the probability to accumulate the DM~\cite{Brayeur}.
Besides, neutron star could capture enough DM due to its high
density~\cite{Bromley}. Though the DM is usually assumed to be
collisionless, the assumption of self-interacting DM can  not be
simply ruled out~\cite{sper00}. Even though the majority of DM is
regarded to be cold and collisionless, Fan et. al. proposed most
recently that a subdominant fraction of DM could have much stronger
interactions~\cite{fan13}. In the following, we consider DM that
features interactions.

\begin{figure}[thb]
\begin{center}
\vspace*{-5mm}
\includegraphics[height=8.0cm,width=10.0cm]{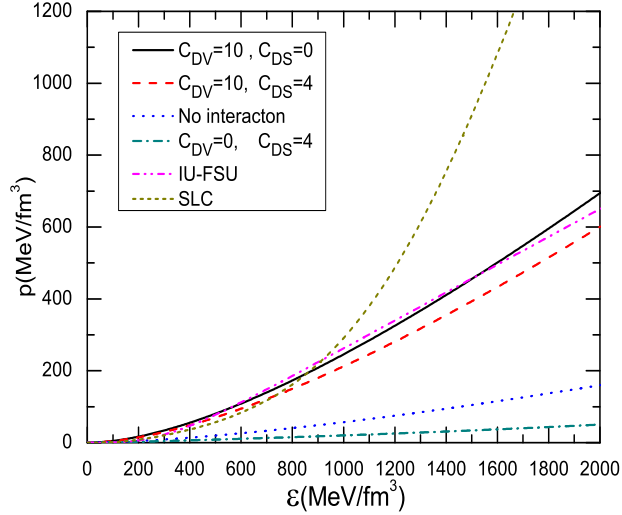}
 \end{center}
\vspace*{-5mm}\caption{(Color online) Pressure and energy density
relationship of DM. The interactions with various strengthes  are
considered. The units of $c_{DV}$ and $c_{DS}$ are $GeV^{-1}$. For
comparisons, the results of NM with the SLC and IU-FSU are also
displayed.\label{EPD}}
\end{figure}

Now, we examine the EOS with various interactions for DM. From
Eqs.(\ref{eq:ed}) and (\ref{eq:pd}), we know that the repulsion
(attraction) potential is just determined by the ratio parameter
$C_{DV}=g_v/m_v$ $(C_{DS}=g_s/m_s)$ in the RMF approximation. Thus,
these ratio parameters can represent the interaction strength. In
order to force the potential in Eq.(\ref{potential}) attractive at
large separations and repulsive at short distances, $m_v$ must be
greater than $m_s$, and $g_v$ must be greater than $g_s$. However,
there are no limits on the size of $C_{DV}$ and $C_{DS}$. Note that
$C_{DS}$ should not be too large. Otherwise, the pressure may become
negative and then be not a monotonous function of energy density,
invalidating the use of the TOV equation. The strengthes we select
here are a little arbitrary, but this does not keep us from drawing a
rough conclusion concerning the DM interaction. Shown in
Fig.~\ref{EPD} is the pressure-energy density relationship of DM.
Here, the free DM particle mass is taken as 1GeV, and we choose
$C_{DV} = 10 GeV^{-1}$ and/or $C_{DV} = 4 GeV^{-1}$ in the
calculation. It is seen in Fig.~\ref{EPD} that the repulsion with a
large $C_{DV}$ stiffens clearly the EOS of DM, while attraction just
softens moderately the EOS of DM because of a much smaller $C_{DS}$.
For comparisons, we also depict the NM results with the SLC and
IU-FSU in Fig.~\ref{EPD}.

\begin{figure}[thb]
\begin{center}
\vspace*{-5mm}
\includegraphics[height=8.0cm,width=14.0cm]{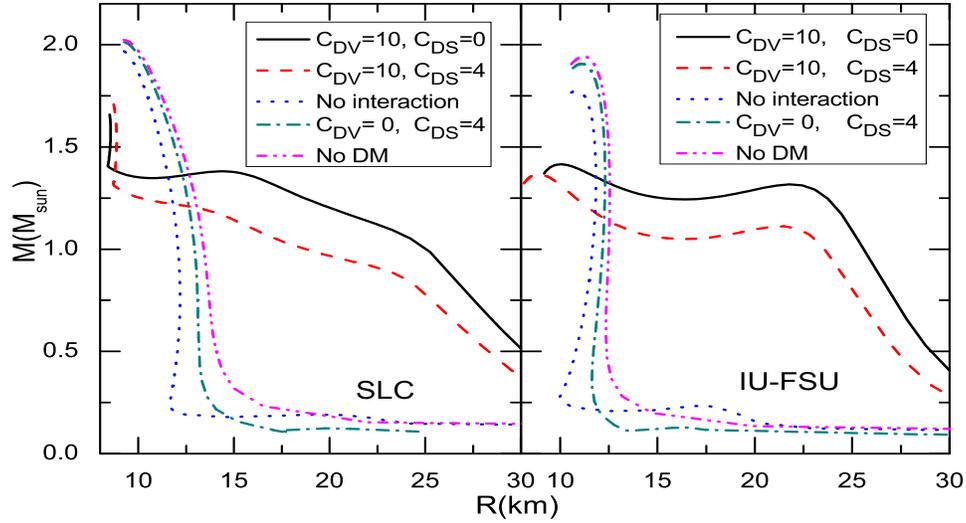}
 \end{center}
\vspace*{-5mm}\caption{(Color online) Mass-radius relations with
various DM EOSs. The attractive and/or repulsive strengthes are
labelled in the legend. \label{MRI}}
\end{figure}

With various DM EOSs, we display the mass-radius relation  in
Fig.\ref{MRI}. Because the repulsion provides resistance to the
gravitational attraction, in principle, the more the repulsion, the
heavier the star. In the one-fluid model, this is simply right.
However, in the two-fluid model, the mass-radius relation is not so
simple. Because of the gravity provided by the other fluid, the
maximum mass of each fluidic constituent reduces, while the repulsion
added by the DM just increases the maximum mass of the DM
constituent. This can be clearly observed by comparing the results of
solid and red-dashed curves in Fig.~\ref{MRI}. In the case of the
pure attraction for DM, we see that the mass-radius relation is just
modified moderately, because the attraction diminishes the DM
proportion in the star. It's obvious that the stiffer the EOS of DM
shown in Fig.~\ref{EPD}, the more effect it could have on the
mass-radius relation of DANSs. Though the maximum mass of neutron
stars is here less than 2$M_\odot$ with the given repulsion for DM,
it can exceed 2$M_\odot$ for stronger repulsion. Note that though
these statements are made from the results obtained with the specific
mass of DM candidate, it is qualitatively valid for other choices of
the DM candidate mass.

\begin{figure}[thb]
\begin{center}
\vspace*{-5mm}
\includegraphics[height=8.0cm,width=14.0cm]{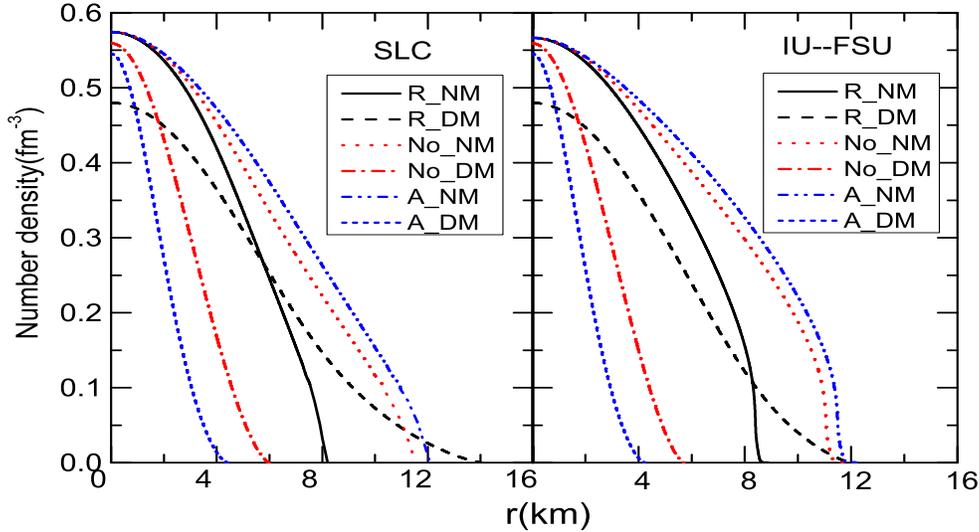}
 \end{center}
\vspace*{-5mm}\caption{(Color online) Number density profiles in
DANSs. In the legend, $R$ ($A$) represents the pure repulsion
(attraction) with the interaction strength $C_{DV}=10 GeV^{-1},
C_{DS}=0 $($C_{DV}=0 , C_{DS}=4GeV^{-1}$), while $No$ represents no
interaction for DM. These cases are corresponding to those in
Figs.\ref{EPD} and \ref{MRI}. \label{HALO}}
\end{figure}

While the interaction of DM leads to various mass-radius relations,
it is interesting to investigate constituent density profiles in
DANSs. Shown in Fig.~\ref{HALO} are various particle number density
profiles for the DANSs with the DM interactions corresponding to
those in Figs.~\ref{EPD} and \ref{MRI}. Here, the central energy
density is fixed to be $600MeV/fm^{3}$ for both NM and DM. In this
case, the mass of the DANS ranges from $1.13 M_{\odot}$ to $1.38
M_{\odot}$ and from $1.34 M_{\odot}$ to$1.60 M_{\odot}$ for the SLC
and IU-FSU, respectively. We can observe  that the attraction of DM
shrinks the DM density distribution, leading to the increase of the
number density gradient of DM in the DANS, together with the decrease
of the number density gradient of NM. The repulsion supports an
spatial extension, opposite to the shrinkage provided by the
attraction. The DM extension out of NM actually generates a clear DM
halo. It is seen from Fig.~\ref{HALO} that the number density of the
DM halo is much larger than that shown in Fig.\ref{NB}. Our
calculation also indicates that it's more likely to form the DM halo
with a lower central energy density. The halo structure is somehow
dependent on nuclear EOS, as shown in Fig.~\ref{HALO}. This is
because various EOSs provide different NM distribution that screens
DM.  Though DM halo is invisible, the visible size of DANSs is now
certainly affected by the interaction of DM. We can address that the
visible radius of DANSs is not only affected by the DM proportion,
but also affected by the interaction of DM. This would affect the
extraction of EOS from astrophysical observations.

\begin{table}[h]
\caption{The allowed amount of the DM accreted onto the normal
neutron star  due to the equilibrium between the gravity and matter
pressure.  The normal neutron star mass is fixed to be $1.30M_\odot$.
$\vep_c$ is the central energy density which is equal for DM and NM,
and $R_{DM}$ $(R_{NM})$ and $M_{DM}/M_\odot$ ($M_{NM}/M_\odot$) are
the respective radius and mass in DANSs. The energy density and
radius are in units of $MeV/fm^{-3}$ and $km$, respectively. The
interaction type in the first column stands for the various
interaction strengths of DM that are the same as the corresponding
cases in Fig.\ref{HALO}.\label{t:accu}}
 \begin{center}
    \begin{tabular}{ l| l c c c c c c }
\hline\hline
&  Model & $\vep_c $  &  $R_{DM}$  &  $M_{DM}/M_\odot$  &  $R_{NM}$  &  $M_{NM}/M_\odot$  &  $M_{SM}/M_\odot$   \\
     \hline
        &SLC   &508    &-    & -    &12.85   &1.30   &1.30   \\
No DM   &SLCd  &621    &-    & -    &11.75   &1.30   &1.30
\\\hline
        &SLC   &620    &4.56 &0.03  &12.40   &1.30   &1.33   \\
Attraction   &SLCd  &725    &4.24 &0.03  &11.39   &1.30   &1.33    \\ \hline
                 &SLC   &820    &5.36    & 0.10    &11.46   &1.30   &1.40   \\
No interaction   &SLCd  &900    &5.12    & 0.10    &10.67   &1.30
&1.40
\\ \hline
             &SLC   &1800    &6.26    & 0.34    &8.53   &1.30   &1.64   \\
Repulsion    &SLCd  &1776    &6.25    & 0.33    &8.31   &1.30   &1.63   \\
\hline\hline
\end{tabular}
\end{center}
\end{table}

The DANS mass varies with the interaction of DM at given central
energy density. By varying the interaction of DM, we may obtain the
different DM proportion in the DANS according to the mechanical
equilibrium between the gravity and matter pressure. Specifically, we
consider the example that the mass from constituent NM is fixed to be
$1.30M_\odot$. The results are tabulated  in Table~\ref{t:accu}.
Here, the interaction strengthes of DM are the same as the
corresponding cases in Fig.~\ref{HALO}, and the central energy
density of NM and DM is set to be equal. From Table~\ref{t:accu}, we
see that the DM amount in the neutron star, allowed by the mechanical
equilibrium, causes the increase of the $\vep_c$ of the star. The
stiffer the EOS of DM, the greater the $\vep_c$. Indeed, the allowed
DM amount accumulated in the DANS is the maximum DM amount beyond
which the DANS should collapse into the black hole. The observation
from Table~\ref{t:accu} indicates that the  maximum mass of the DANS
increases surely by changing the interaction from the attraction to
repulsion. We may thus speculate that the DANS maximum mass further
increases to exceed the 2$M_\odot$ restriction~\cite{De10,An13} by
increasing the repulsion strength. This is the case that can be
verified numerically. Moreover, it is interesting to point out that
the excess of 2$M_\odot$ can be realized by increasing the ratio of
the DM central energy density to that of NM  without changing the
repulsion strength used in Table~\ref{t:accu}. Nevertheless, as the
DM amount in the DANS increases to provide more gravitation, NM in
the DANS is more compressed and becomes more compact.

\begin{figure}[thb]
\begin{center}
\vspace*{-5mm}
\includegraphics[height=12.0cm,width=14.0cm]{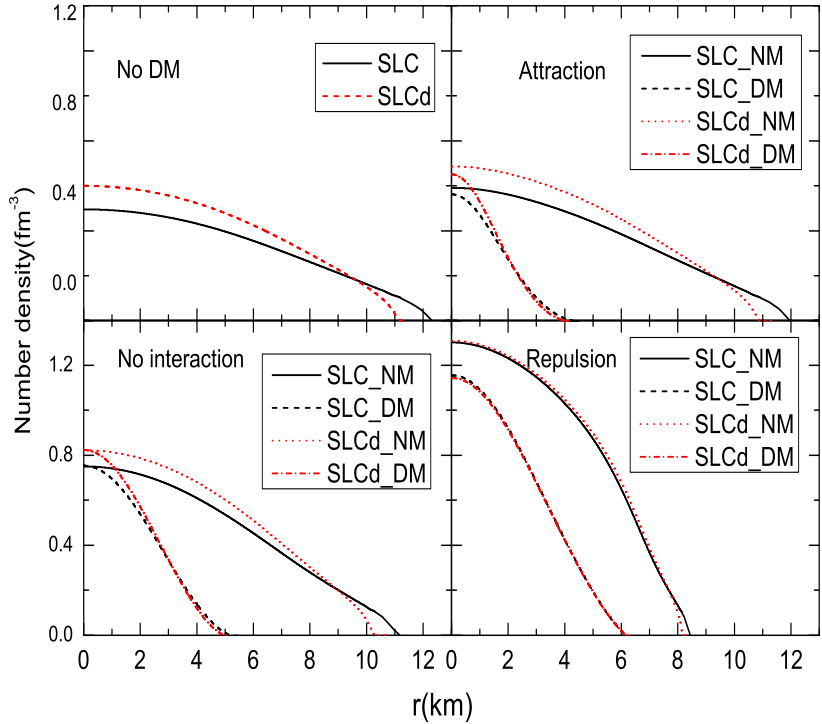}
 \end{center}
\vspace*{-5mm}\caption{(Color online) Number density profiles in the
DANS with different DM interactions for models SLC and
SLCd.\label{NM2}}
\end{figure}

In Fig.~\ref{NM2}, we plot various number density profiles for
neutron stars with the masses and corresponding interactions same as
those in Table~\ref{t:accu}. In the present case, we see that DM
resides in the central region of DANS and no DM halo forms. We see in
Fig.~\ref{NM2} that  DM in the DANS can influence the distribution of
NM. In the case without DM, we know that the different matter
distribution with the SLC and SLCd is attributed to their different
nuclear symmetry energy. It is interesting to see that DM in the DANS
results in the decrease of the difference in the NM density profiles.
Moreover, the decrease can be enhanced by stiffening the DM EOS. If
appropriate repulsion is assumed to have between DM, the difference
in density profiles and consequently in star radii given by the SLC
and SLCd can run almost to disappear. In this case, the difference in
neutron star radii turns out to be quite small, as seen in
Table~\ref{t:accu}. Though the models  are different just in the
density dependence of nuclear symmetry energy, the astrophysical
extraction of the constraint on the symmetry energy seems to be
difficult if the neutron star is admixed with DM featuring
sufficiently strong repulsion.

\section{Summary}
\label{summary}

In this work, we have studied the effects of fermionic DM on the
properties of neutron stars using the two-fluid TOV formalism. While
by assuming that the DM and NM have  no interactions but the
gravitation, we have considered various choices for masses  and
interactions for DM candidates. It is found that the mass-radius
relationship of DANSs depends sensitively on the mass of DM
candidate, the amount of DM, and interactions among DM candidates.
The existence of DM in DANSs results in a spread of mass-radius
relationships that cannot be interpreted with a unique equilibrium
sequence. The mass region of DM candidates where DM affects
significantly the DANS properties is found to be below a few GeV. An
important consequence of the DM admixture in neutron stars  is the
shrinkage of the NM surface that actually results in the small radius
observation. Interestingly, we find that the difference in density
profiles in the DANS and consequently in star radii caused by various
density dependencies of nuclear symmetry energy tends to disappear,
as long as the repulsion of accumulated DM is sufficient. It is
interesting to see that the DM distribution can surpass the NM
distribution to form DM halo especially for the DM candidates with
the low mass. Generally, an explicit DM halo may arise favorably if
the repulsion of DM exists. These phenomena indicate that the
admixture of DM in neutron stars can significantly affect the
astrophysical extraction of nuclear EOS through neutron star
measurements. Provided the nuclear EOS can be well determined in the
terrestrial laboratory, it would be hopeful to look for the evidence
for DM using the phenomena that are sensitive to the DM admixture in
the neutron star observations. On the other hand, the DM admixture
can affect the maximum mass of neutron stars. Though the maximum mass
tends to be reduced by the DM accretion, the practical situation is
not so simple and actually depends on very details of models, the DM
fraction and interactions. In particular, favorably with the large DM
fraction in the DANS the repulsive interaction among DM can lead to
massive stars with the mass above 2$M_\odot$.

\section*{Acknowledgement}

The work was supported in part by the SRTP Grant of the Educational
Ministry No. 1210286047,  the National Natural Science Foundation of
China under Grant Nos. 10975033 and 11275048,  the China Jiangsu
Provincial Natural Science Foundation under Grant No.BK20131286, and
the China Scholarship Council.

\appendix
\section{Derivation of  Eq.(\ref{eq:TOVS})}
\label{append}

While one can find the derivation of the one-fluid TOV equation in
the literature, we give a detailed deduction of the two-fluid TOV
equations [Eq.(\ref{eq:TOVS})] according to the stellar stability
condition~\cite{wei72}. With the uniform entropy per particle and
chemical composition for both normal matter and dark matter, the
equilibrium of a particular stellar configuration reaches, if and
only if the $M$ defined by
\begin{equation}
 M= \int_0^{ \infty} 4 \pi r^2 [\vep_N (r) + \vep_D (r) ]dr,
\end{equation}
is stationary with respect to all variations of $\vep_N(r)$ and
$\vep_D(r)$ that maintain the conservation of the two qualities
\begin{equation}
N_D=\int_0^{\infty} 4\pi r^2 \rho_D(r) [1-\frac{2GM(r)}{r} ]^{-1/2}
dr,
\end{equation}
and
\begin{equation}
N_N=\int_0^{\infty} 4\pi r^2 \rho_N(r) [1-\frac{2GM(r)}{r} ]^{-1/2}
dr,
\end{equation}
and that leave the entropy per particle and the chemical composition
unchanged.

We introduce two Lagrange multipliers $\lambda$ and $\alpha$ to
demonstrate the fact that $M$ will be stationary with respect to all
variations that leave $N_N$ and $N_D$ fixed if and only if  $ \delta
M -\lambda \delta N_N -\alpha \delta N_D$ is stationary with respect
to all variations. The functional variation for the given variation
$\delta \vep_N (r) $ and $\delta \vep_D (r) $ can be written as
\begin{eqnarray}\label{eq:detm}
   &&  \delta M -\lambda \delta N_N -\alpha \delta N_D  \nonumber\\
 & = &
     \int_0^{\infty} 4\pi r^2 [\delta \vep_N(r) +\delta \vep_D(r)]dr  \nonumber\\
&-&\lambda \int_0^{\infty} 4\pi r^2 [1-\frac{2GM(r)}{r} ]^{-1/2}
\delta \rho_N(r)dr\\
 &-&\lambda G \int_0^{\infty} 4\pi r
[1-\frac{2GM(r)}{r} ]^{-3/2} \rho_N(r) \delta M(r)dr \nonumber\\
&-&\alpha \int_0^{\infty} 4\pi r^2 [1-\frac{2GM(r)}{r} ]^{-1/2}
\delta \rho_D(r)dr \nonumber \\
 &-&\alpha G \int_0^{\infty} 4\pi r
[1-\frac{2GM(r)}{r} ]^{-3/2} \rho_D(r) \delta M(r)dr \nonumber,
   \end{eqnarray}
with
\begin{equation}\label{eq:mr}
\delta M(r)= \int_0^r 4 \pi {r^{\prime}}^2 [ \delta \vep_N
(r^{\prime}) +\delta \vep_D (r^{\prime}) ]dr^{\prime}.
\end{equation}
Note that the integrands vanish outside $R+\delta R$ and we use the
upper limit $\infty$ just for convenience.  These variations are
supposed not to change the entropy per nucleon and DM particle,
namely
\begin{eqnarray}
 \delta (\frac{\vep_N (r)}{\rho_N(r)}) + p_N (r) \delta (\frac{1}{\rho_N(r)})  &=& 0 , \\
 \delta (\frac{\vep_D (r)}{\rho_D(r)}) + p_D (r) \delta
(\frac{1}{\rho_D(r)}) &=&0,
\end{eqnarray}
and thus
\begin{eqnarray}
\nonumber
\delta \rho_N(r) =\frac{\rho_N (r)}{p_N (r) +\vep_N (r)} \delta \vep_N (r), \\
\delta \rho_D(r) =\frac{\rho_D (r)}{p_D (r) +\vep_D (r)} \delta
\vep_D (r).
\end{eqnarray}

By substituting Eq.(\ref{eq:mr}) into Eq.(\ref{eq:detm}) and
interchanging the $r$ and $r^\prime$ integrals,  we have
 \begin{eqnarray}
 && \delta M -\lambda \delta N_N -\alpha \delta N_D  \nonumber\\
   &=&
     \int_0^{\infty} 4\pi r^2 \{ 1-\frac{\lambda \rho_N (r)}{p_N(r) +\vep_N (r)}
     [1-\frac{2GM(r)}{r}]^{-1/2}\nonumber \\
   && - \lambda F_N(r)- \alpha F_D(r) \}\delta\vep_N dr \nonumber\\
&+&     \int_0^{\infty} 4\pi r^2 \{ 1-\frac{\alpha \rho_D (r)}{p_D(r)
+\vep_D (r)} [1-\frac{2GM(r)}{r}]^{-1/2} \\
     && - \lambda F_N(r)- \alpha F_D(r)  \}\delta\vep_D dr,\nonumber
\end{eqnarray}
where we have defined two quantities
\begin{eqnarray}
   \nonumber
 F_N(r) &=&  G \int_r^{\infty} 4\pi r^{\prime} \rho_N (r^{\prime} )dr^{\prime}
[1-\frac{2GM(r^{\prime} )}{r^{\prime} }]^{-3/2} dr^{\prime}, \nonumber \\
F_D(r) &=&  G \int_r^{\infty} 4\pi r^{\prime} \rho_D (r^{\prime})
dr^{\prime} [1-\frac{2GM(r^{\prime} )}{r^{\prime} }]^{-3/2}
dr^{\prime}.\nonumber
\end{eqnarray}
We see that  $\delta M -\lambda \delta N_N -\alpha \delta N_D  $ will
vanish for all  $\delta \vep_D (r) $ and  $\delta \vep_N (r)$ if and
only if
\begin{eqnarray}
1-\frac{\lambda \rho_N (r)}{p_N(r) +\vep_N (r)}
[1-\frac{2GM(r)}{r}]^{-1/2}
- \lambda F_N -\alpha F_D &=& 0,  \label{la1}\\
1-\frac{\alpha \rho_D (r)}{p_D(r) +\vep_D (r)}
[1-\frac{2GM(r)}{r}]^{-1/2} - \lambda F_N -\alpha F_D &=&0,
\label{la2}
\end{eqnarray}
while two equations can be rewritten as
\begin{eqnarray}
\frac{ \rho_N (r)}{p_N(r) +\vep_N (r)} [1-\frac{2GM(r)}{r}]^{-1/2}
+F_N +  \frac{\alpha}{\lambda} F_D
 &=&\frac{1}{\lambda},\label{eq:dn1}
 \\
\frac{ \rho_D (r)}{p_D(r) +\vep_D (r)} [1-\frac{2GM(r)}{r}]^{-1/2}
+F_D +  \frac{\lambda}{\alpha} F_N &=&\frac{1}{\alpha},\label{eq:dn2}
\end{eqnarray}
with
\begin{equation}
\frac{\lambda}{\alpha}=\frac{\rho_D(r) [p_N (r)+\vep_N
(r)]}{\rho_N(r) [p_D (r)+\vep_D (r)]},
\end{equation}
obtained by equating the left-hand sides of  Eqs.(\ref{la1}) and
({\ref{la2}). Because the right-hand side of Eqs.(\ref{eq:dn1}) and
(\ref{eq:dn2}) is independent of the coordinate $r$, the
differentiation over the coordinate vanishes. By differentiating both
sides of Eq.(\ref{eq:dn1}), we have
\begin{eqnarray} \label{N11}
\nonumber &&\{\frac{\rho_N^\prime (r)}{p_N
(r)+\vep_N(r)}-\frac{\rho_N(r) [p_N^\prime (r)+\vep_N^\prime
(r)]}{[p_N (r)+\vep_N(r)]^2}\} [1-\frac{2GM(r)}{r}]^{-1/2} \\
\nonumber && +\frac{  G \rho_N (r)}{p_N(r) +\vep_N (r)} \{4\pi r
[\vep_D (r) +\vep_N (r)]-\frac{M(r)}{r^2} \}
[1-\frac{2GM(r)}{r}]^{-3/2}  \\ \nonumber
&& - 4\pi r G \rho_N (r) [1-\frac{2GM(r)}{r}]^{-3/2} \\
&& - 4\pi r G \rho_N (r) \frac{p_D(r)+\vep_D(r)}{p_N(r)+\vep_N(r)}
[1-\frac{2GM(r)}{r}]^{-3/2}=0
\end{eqnarray}
The condition of uniform entropy per nucleon gives
\[
\frac{d}{dr}(\frac{\vep_N (r)}{\rho_N (r)}) +p_N
\frac{d}{dr}(\frac{1}{\rho_N(r)})=0,
\]
which leads to the relation
\begin{equation}\label{eq:en1}
   \rho_N^\prime (r) = \frac{\rho_N(r) \vep_N^\prime (r)}{p_N(r) + \vep_N
   (r)}.
\end{equation}
Substituting Eq.(\ref{eq:en1}) into Eq.(\ref{N11}), we get the TOV
equation for normal matter
\begin{equation}\label{eq:tovn}
   p_N^\prime (r) =  -G[p_N(r)+\vep_N(r)]
   [1-\frac{2GM(r)}{r}]^{-1}\{ 4\pi r [p_N(r)+p_D(r)] +\frac{M(r)}{r^2} \}
\end{equation}
Similarly, we can obtain the TOV equation for dark matter as:
\begin{equation}\label{eq:tovd}
   p_D^\prime (r) =  -G[p_D(r)+\vep_D(r)]
   [1-\frac{2GM(r)}{r}]^{-1}\{ 4\pi r [p_N(r)+p_D (r)] +\frac{M(r)}{r^2} \}
\end{equation}
Eqs.(\ref{eq:tovn}) and (\ref{eq:tovd}) are exactly the TOV equations
given in Eq.(\ref{eq:TOVS}).

\end{document}